# On-Chip Implementation of Pipeline Digit-Slicing Multiplier-Less Butterfly for Fast Fourier Transform Architecture

[1]Yazan Samir Algnabi, [1,2]Rozita Teymourzadeh, [1]Masuri Othman, [1]Md Shabiul Islam
[2]Mok Vee Hong

[1]Institute of MicroEngineering and Nanoelectronics IMEN, VLSI Design Department,
Universiti Kebangsaan Malaysia, 43600 Bangi, Selangor, Malaysia
[2]Faculty of Engineering, Architecture and Built Environment,
Electrical & Electronic Engineering department, UCSI University, Kuala Lumpur, Malaysia

**Abstract:** The need for wireless communication has driven the communication systems to high performance. However, the main bottleneck that affects the communication capability is the Fast Fourier Transform (FFT), which is the core of most modulators. This study presents on-chip implementation of pipeline digit-slicing multiplier-less butterfly for FFT structure. The approach taken; in order to reduce computation complexity in butterfly, digit-slicing multiplier-less single constant technique was utilized in the critical path of Radix-2 Decimation In Time (DIT) FFT structure. The proposed design focused on the trade-off between the speed and active silicon area for the chip implementation. The new architecture was investigated and simulated with MATLAB software. The Verilog HDL code in Xilinx ISE environment was derived to describe the FFT Butterfly functionality and was downloaded to Virtex II FPGA board. Consequently, the Virtex-II FG456 Proto board was used to implement and test the design on the real hardware. As a result, from the findings, the synthesis report indicates the maximum clock frequency of 549.75 MHz with the total equivalent gate count of 31,159 is a marked and significant improvement over Radix 2 FFT butterfly. In comparison with the conventional butterfly architecture, design that can only run at a maximum clock frequency of 198.987 MHz and the conventional multiplier can only run at a maximum clock frequency of 220.160 MHz, the proposed system exhibits better results. The resulting maximum clock frequency increases by about 276.28% for the FFT butterfly and about 277.06% for the multiplier. It can be concluded that on-chip implementation of pipeline digit-slicing multiplier-less butterfly for FFT structure is an enabler in solving problems that affect communications capability in FFT and possesses huge potentials for future related works and research areas.

## INTRODUCTION

FFT plays an important role in many Digital Signals Processing (DSP) applications such as in communication systems and image processing. It is an efficient algorithm to compute the Discrete Fourier Transform (DFT). DFT is the main and important procedure in data analysis, system design, and implementation (Oppenheim and Rader, 1990). In order to reduce the complexity computation of the FFT algorithm many modules have been designed and implemented in different platforms. These modules focus on the radix order or twiddle factors to perform a simple and efficient algorithm which includes the higher radix FFT (Bergland, 1969), the mixed-radix FFT (Singleton, 1969), the prime-factor FFT (Kolba and Parks, 1977), the recursive FFT (Varkonyi-Koczy, 1995), low-memory reference FFT (Wang *et al*., 2007), Multiplier-less based FFT (Zhou *et al*., 2007; Prasanthi *et al*., 2005; Mahmud and Othman, 2006) and Application-Specific Integrated Circuits (ASIC) system such as stated by Baas (1999). ASIC-based systems are able to fit real low-power or high performance applications; however the function is very solid to be modified (Hsu and Lin, 2008). The study of the digit-slicing technique has been dealt by Bin Nun and Woodward (1976); Peled and Liu (1976); and Sharrif, (1980) for the digital filters.

**Corresponding Author:** Rozita Teymourzadeh, Institute of Microengineering and Nanoelectronics IMEN,
VLSI Design Department, Universiti Kebangsaan Malaysia, 43600 Bangi, Selangor, Malaysia

The design and implementation of Digit-slicing FFT has been discussed by Samad *et al.*, (1998). This study proposed a similar idea with the ones put forth by Samad *et al* (1998); but having a difference by the use of a different algorithm and different platform, which helps to improve the performance and achieve higher speed. Recently, FPGAs Field Programmable Gate Array have become an applicable option to direct hardware solution performance in the real time application. In this study, digit-slicing architecture was proposed in designing the pipeline digit-slicing multiplier-less butterfly. The FFT butterfly multiplication is the most crucial part in causing the delay in the computation of the FFT. In view of the fact, the twiddle factors in the FFT processor were known in advance hence we proposed to use the pipeline digit slicing multiplier-less butterfly to replace the traditional butterfly in FFT.

The study structure is organized as follows; describes the FFT architecture in brief, explains the butterfly conventional architecture, discuses the digit slicing architecture, explicates the design of the pipeline digit- slicing multiplier-less butterfly architecture in detail and finally the implementation result and conclusion respectively.

**Fast Fourier Transform (FFT):** A useful method to transform domains from the time domain to the frequency domain and the reverse for the implementation on digital hardware is the DFT. For N-point DFT of a complex data sequence x (n) is defined in Eq. 1:

$$X(k) = \sum_{n=0}^{N-1} x(n) W_N^{kn}, k = 0,1,......, N-1 \qquad (1)$$

Where:
x(n) and X(k) = Complex numbers
$W_N^{kn} = e^{-j2\pi/N}$ = The twiddle factor

The DFT of N-point finite sequence represents harmonically related frequency components of x(n). The direct computation of Eq. 1 requires the order of $N^2$ operations where N is the transform size. Cooley and Tukey (1965) found this new technique to reduce the order of complexity operations of DFT from $N^2$ to (Nlog$_2$N). Consequently, a huge number of FFT algorithms have been developed such as Radix-2, radix-4 and split radix algorithms. These algorithms are mostly used for practical applications due to their simple structure and constant butterfly geometry.

In general, higher-radix FFT algorithm has fewer numbers of complex multiplications, whereas radix-2 FFT algorithm is the simplest form in all FFT algorithms. Furthermore, it has a regularity mode that makes it suitable for VLSI implementation as shown in the fallowing Eq. 2:

$$X[m] = \sum_{n=0}^{\frac{N}{2}-1} x[2n] W_{\frac{N}{2}}^{nm} + W_N^m \sum_{n=0}^{\frac{N}{2}-1} x[2n+1] W_{\frac{N}{2}}^{nm} \qquad (2)$$

FFT algorithm relies on a 'divide-and-conquer' methodology, which divides the N coefficient points into smaller blocks in different stages. The first stage computes with groups of two coefficients, yielding N/2 blocks, each computing the addition and subtraction of the coefficients scaled by the corresponding twiddle factors, called a butterfly for its cross-over appearance as shown in Fig. 1.

These results are used to compute the next state of N/4 blocks, which will then combine the results of two previous blocks, combining four coefficients at this point. This process is repeated until one main block is formed, with a final computation of all N coefficients. Fig. 2 shows the 8-point radix-2 DIT FFT.

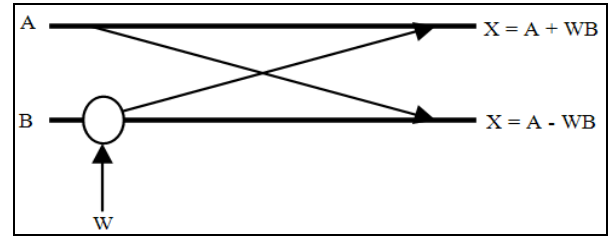

Fig. 1: Butterfly structure.

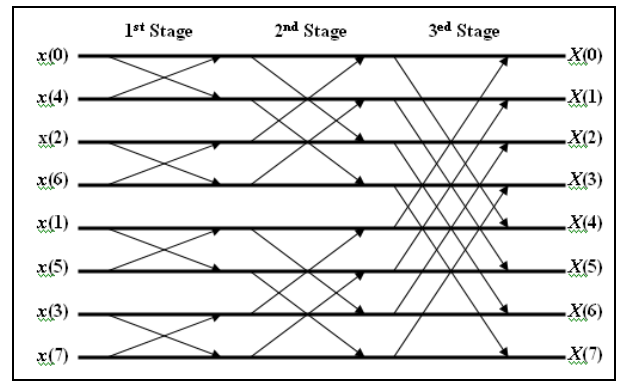

Fig. 2: 8-points FFT radix-2 Decimation in Time.

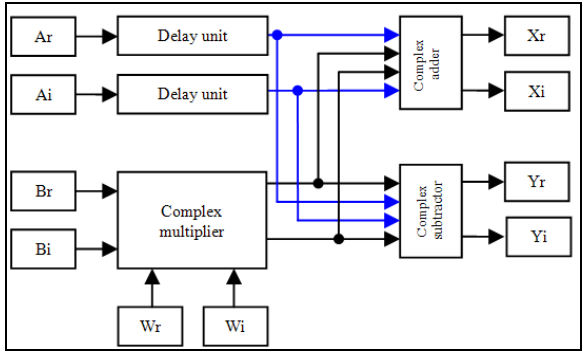

Fig. 3: Radix-2 DIT FFT Butterfly Architecture.

**Conventional Butterfly architecture** The conventional radix-2 DIT butterfly architecture consists of complex data I/O, complex multiplier and complex adder and subtraction as shown in Fig. 3.

Consider A and B as the complex input data, and the complex twiddle factor is considered as $W = Wr-jWi$, hence finally the complex output are X and Y.

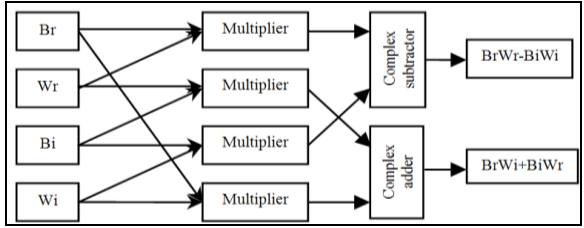

Fig. 4: Complex multiplier structure

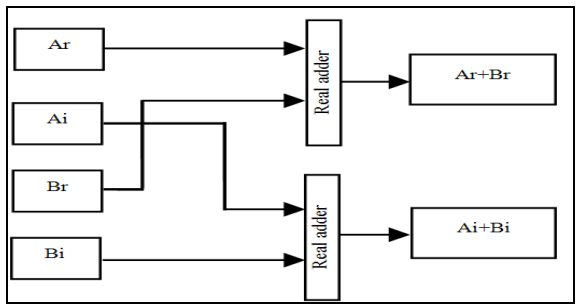

Fig. 5: Complex Adder Structure

The index r and i represent the real and imaginary parts respectively:

$$X = A + WB \quad (3)$$
$$Y = A - WB \quad (4)$$

$$(Xr + jXi) = (Ar + jAi) + \left[(Wr + jWi) \times (Br + jBi)\right] \quad (5)$$
$$(Yr + jYi) = (Ar + jAi) - \left[(Wr + jWi) \times (Br + jBi)\right] \quad (6)$$

The implementation of the complex multiplier is required for four real multipliers and two real adders as shown in Fig. 4. The complex multiplier is determined in Eq.7:

$$\begin{aligned}(Br + jBi) \times (Wr + jWi) &= (Br \times Wr) + (Br \times jWi) \\ &\quad + (jBi \times Wr) + (jBi \times jWi) \\ &= [(Br \times Wr) + (jBi \times jWi)] \\ &\quad + [(Br \times jWi) + (jBi \times Wr)] \\ &= [(Br \times Wr) - (Bi \times Wi)] \\ &\quad + [(Br \times jWi) + (jBi \times Wr)]\end{aligned} \quad (7)$$

The real and imaginary parts of the multiplication result is $[(Br \times Wr) - (Bi \times Wi)]$ and $[(Br \times jWi) + (jBi \times Wr)]$ respectively.

The complex adder is required for two real adders to perform addition functionality as shown in Fig. 5.

$$(Ar + jAi) + (Br + jBi) = (Ar + Br) + j(Ai + Bi) \quad (8)$$

**Digit-slicing architecture:** The concept behind the digit-slicing architecture is any binary number that can be sliced into a few blocks of shorter binary numbers, with each block carrying a different weight. In this study, the fixed-point 2's complements arithmetic has been chosen to represent the input data, which are singed numbers with absolute value less than one. The absolute value of the input data x with length of B bits $(x^0, x^1, x^2, \ldots, x^{B-1})$ has been represented in 2's complement as:

$$x = \sum_{k=0}^{B-1} 2^{-j} x^j \quad (9)$$

To represent the sliced data, there are many different algorithms. Depending on the data type and word length, different structures can be introduced. In this study, where the fundamental sliced algorithm will be presented as following:

$$x = \left[\sum_{k=0}^{b-1} 2^{pk} X_k\right] 2^{-(pb-1)} \quad (10)$$

Where:
  x = Sliced into b blocks
  p = Bit widths per block

$$X_k = \sum_{j=0}^{p-1} 2^j X_{k,j} \quad (11)$$

Where:
$X_{k,j}$ = All either ones or zeros except

$X_{k=b-1, j=p-1}$ = which is zero or minus one

The algorithm in Eq. (10) applies when the sliced data word length is $2^x$ such as $2^2=4$, $2^3=8$, 16… bits. Thus, let us consider the decimal number -0.65625 of which we would like to demonstrate how digit slicing operates accordingly:

$x = 1.010\ 1100_2 = -0.65625_{10}$

where, the suffix 2 refers to a binary fixed point two's complement number 8 bits and the suffix 10 refers to a decimal number, if x is sliced into two blocks, of each four bits wide, that is b = 2 and p = 4:

$$X_0 = \sum_{j=0}^{3} 2^j X_{0,j} = 2^3 + 2^2 = 12$$

$$X_1 = \sum_{j=0}^{3} 2^j X_{1,j} = -2^3 + 2^1 = -6$$

$$x = \left[\sum_{k=0}^{1} 2^{4k} X_k\right] 2^{-(8-1)}$$

$$x = \left[2^{4\times 0} \times 12 + 2^{4\times 1}(-6)\right] \times 2^{-7}$$

$$x = (12 - 96) \times 2^{-7} = \frac{-84}{128} = -0.65625_{10}$$

Another algorithm that represents the sliced data with a word length $2^x+1$ such as $2^2+1=5$, 9, 17…bits can be dealt as the following:

$$x = \sum_{k=0}^{p-1} \left[2^p\right]^{-k} X_k \quad (12)$$

Where, x is a decimal number whose absolute value is less than one and is sliced into b blocks each of p bits wide.

The most significant block is k = 0 where this contains the only sign bit of x plus leading dummy zeros to make up a block of length p bits (Samad *et al.*, 1998):

$X_{k=0} = 0$ or $-1$ only

$$X_k = \sum_{j=0}^{p-1} 2^j X_{k,j}; \quad X_{k,j} = 0 \text{ or } 1 \text{ only for } k \neq 0 \quad (13)$$

Let us assume that the decimal number - 0.328125 is represented as nine bits two's complement number:

$$x = \sum_{k=0}^{2} \left[2^4\right]^{-k} X_k$$

$$x = [2^4]^{-0}[-1] + [2^4]^{-1}[2^3 + 2^1] + [2^4]^{-2}[2^3 + 2^2]$$

$$= -1 + 2^{-1} + 2^{-3} + 2^{-5} + 2^{-6} = -0.328125_{10}$$

As a comparison between the first and the second algorithms, the second algorithm requires one extra block to deal with the sign bit which makes the design more complicated and requires more hardware for the implementation. In this study, the first digit-slicing algorithm has been chosen to build the digit-slicing FFT butterfly structure. Therefore, any complex numbers, F, can be sliced into smaller blocks b, each having a shorter word length, p, as illustrated in following equations:

$$F = F_R + jF_I \quad (13)$$

$$F = \left[\sum_{k=0}^{b-1} 2^{pk} F_{Rk}\right] 2^{-(pb-1)} + j\left[\sum_{k=0}^{b-1} 2^{pk} F_{Ik}\right] 2^{-(pb-1)} \quad (14)$$

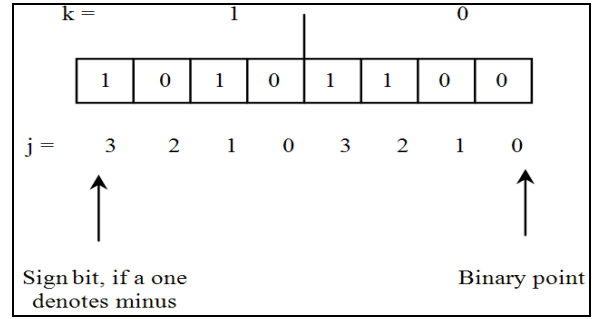

Fig. 6: The digit-slicing first algorithm for -0.65625

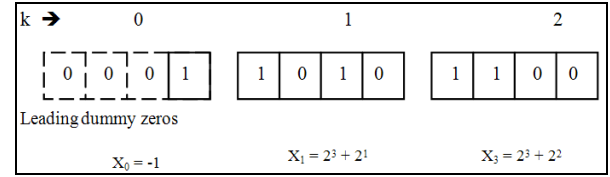

Fig. 7: The digit-slicing 2nd algorithm for -0.328125

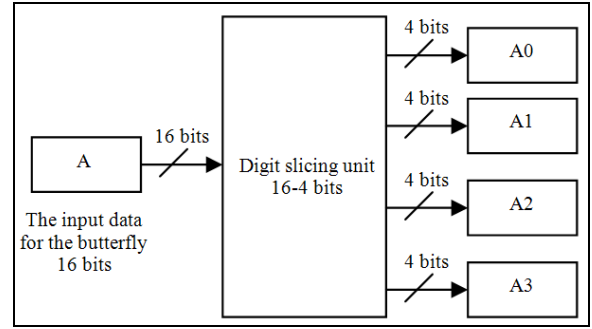

Fig. 8: Digit-slicing structure for the input A.

$$F_{Rk} = \sum_{j=0}^{p-1} 2^j F_{Rk,j} \quad (15)$$

$$F_{Ik} = \sum_{j=0}^{p-1} 2^j F_{Ik,j} \quad (16)$$

Where, the values of $F_{Ik,i}$ and $F_{Rk,I}$ are either zero or one.

**Pipeline digit-slicing multiplier-less butterfly architecture:** The butterfly is the smallest component to build the FFT. As mentioned in the explanations prior to this, the butterfly structure contains one complex multiplier, one complex adder, and one complex subtractor.

The digit-slicing architecture has been applied for the butterfly input to slice the data into four groups each carrying four bits as shown in Fig. 8.

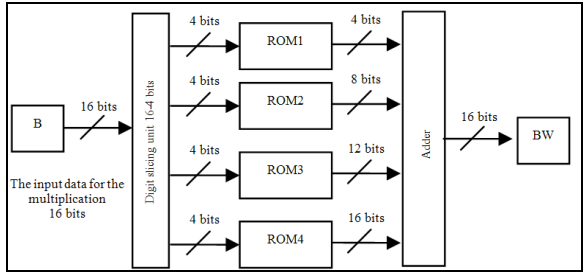

Fig. 9: Digit-Slicing Single Constant Multiplier (DSSCM) Structure.

$$A = \left[\sum_{k=0}^{b-1} 2^{pk} A_k\right] 2^{-(pb-1)} \quad (17)$$

$$A_k = \sum_{j=0}^{p-1} 2^j A_{k,j} \quad (18)$$

where, $A_{k,j}$ are all either ones or zeros except for $A_{k=b-1,j=p-1}$ which is zero or minus one.
The same applies for the input B:

$$B = \left[\sum_{k=0}^{b-1} 2^{pk} B_k\right] 2^{-(pb-1)} \quad (19)$$

$$B_k = \sum_{j=0}^{p-1} 2^j B_{k,j} \quad (20)$$

where, $B_{k,j}$ are all either ones or zeros except for the value $B_{k=b-1,j=p-1}$ which is zero or minus one.

The multiplication functionality is regarded as the most important operation for most signal processing systems, but it is a complex and expensive operation. Many techniques have been introduced for reducing the size and improving the speed of multipliers. Some applications require Constant Coefficient Multipliers such as digital signal processing, image processing, and multiple precision arithmetic in the design of compilers. Constant Coefficient Multipliers is one of the most common solutions to speed up the multiplication process.

The multiplier can be designed for one constant which is termed as Single Constant Multiplier (SCM) or for many constant and is termed as Multiple Constant Multiplier (MCM). Since the twiddle factor in FFT processor are known in advance, a special design of SCM has been proposed to perform the multiplication function with the twiddle factor without using the traditional multiplier, which is termed as Single Constant Multiplier Less (SCML). The design of the SCML consists of four lookup tables (ROMs) and adder to perform the output as shown in Fig 9. To generate the lookup tables data (the multiplication result possibilities), which are 16 different results for each ROM, a special MATLAB program has been written by applying the digit-slicing algorithm for all the possible numbers for the input data (4 bits) from "0000" to "1111" to perform all the possibilities for the multiplication result. The result for the SCML has been optioned by simple addition for all the lookup tables' results. In the hardware implementation, the addition logic has been reduced. During the addition of the four products obtained from the look-up tables, the least significant digit (4 bits) for each level is always added to zero. These bits will not be affected, or changed and will be carried into the next column. The storage of all these possibilities in four different ROMs allows the design to perform the multiplication process without any real multiplier.

From Eq. 10 and 11, the digit-slicing multiplier is represented as the following:

$$BW = \left[\sum_{k=0}^{3} 2^{4k} WB_k\right] 2^{-(7)} \quad (21)$$

$$WB_k = \sum_{j=0}^{3} 2^j WB_{k,j} \quad (22)$$

where, $WB_{k,j}$ are all either ones or zeros except for $WB_{k=b-1,j=p-1}$ which is zero or minus one and where W is the constant.

The result of the multiplication will be added and subtracted with the complex inputs $A_r+jA_i$ for the butterfly to perform the butterfly outputs.

The butterfly output X has been defined as:

$$X = \left[\sum_{k=0}^{b-1} 2^{pk} X_k\right] 2^{-(pb-1)} \quad (23)$$

$$X_k = \sum_{j=0}^{p-1} 2^j X_{k,j} \quad (24)$$

where, $X_{k,j}$ are all either ones or zeros except for $X_{k=b-1,j=p-1}$ which is zero or minus one.

By applying Eq. 17, 19 and 21 into Eq. 3:

$X = A + WB$

$\left[\sum_{k=0}^{3} 2^{4k} X_k\right] 2^{-(pb-1)} = \left[\sum_{k=0}^{3} 2^{4k} A_k\right] 2^{-(pb-1)} +$

$\left[\sum_{k=0}^{3} 2^{4k} WB_k\right] 2^{-(pb-1)}$

$X_k = A_k + WB_k$

$X_k$ is complex number

$X_k = X_{rk} + jX_{ik}$

Real part of $X_{rk} = A_k + WB_{r\ k}$

Imag part of $X_{ik} = A_k + WB_{i\ k}$

The same step for the output X has been applied to get the output Y:

$Y = A - WB$

$\left[\sum_{k=0}^{b-1} 2^{pk} Y_k\right] 2^{-(pb-1)} = \left[\sum_{k=0}^{b-1} 2^{pk} A_k\right] 2^{-(pb-1)} -$

$\left[\sum_{k=0}^{b-1} 2^{pk} WB_k\right] 2^{-(pb-1)} \quad (26)$

$Y_k = A_k - WB_k$

$Y_k$ is complex number

$Y_k = Y_{rk} + jY_{ik}$

Real part of $Y_{rk} = A_k - WB_{r\ k}$    Imag part of $Y_{ik} = A_k - WB_{i\ k}$

Finally, the complex output is represented as the following:

$X_{rk} = A_{rk} + WB_{rk} + WB_{ik}$    (27)
$X_{ik} = A_{ik} + WB_{ik} - WB_{rk}$    (28)
$Y_{rk} = A_{rk} - WB_{rk} - WB_{ik}$    (29)
$Y_{ik} = A_{ik} - WB_{ik} + WB_{rk}$    (30)

The full digit-slicing single constant multiplier-less has been designed and tested in MATLAB as shown in Fig 10 and 11, of which the result is then compared with the normal multiplier.

For the addition and subtraction, the parallel-prefix Koggie and Stone Ling adder were used for high speed and better performance. The pipeline technique was applied for the full design for better performance.

## RESULT

Two different modules were implemented for radix-2 DIT butterfly. The first module uses the conventional architecture for the butterfly where the twiddle factors are stored in ROM and called by the butterfly to be multiplied with the inputs by utilising the dedicated high speed multiplier equipped with the Virtex-II FPGA.

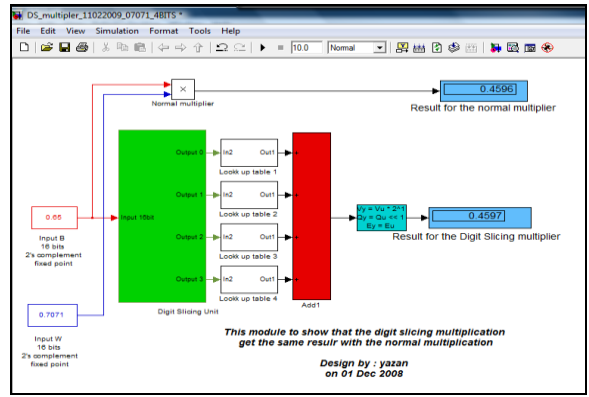

Fig. 10: MATLAB design of Digit-Slicing Single Constant Multiplier-Less for the Butterfly.

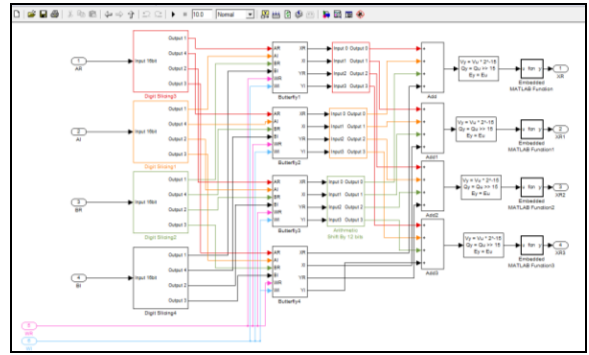

Fig. 11: MATLAB design of Digit-slicing Butterfly.

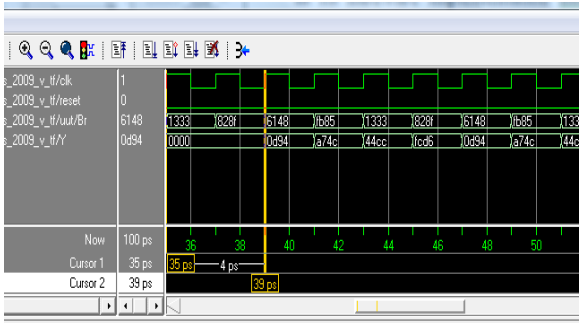

Fig. 12: Simulation result of the Pipeline Digit-slicing Single Constant Multiplier-Less for the Butterfly.

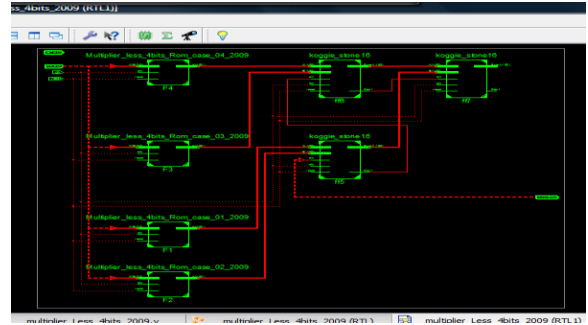

Fig. 14: RTL schematic for the Pipeline Digit-slicing Single Constant Multiplier-Less for the Butterfly.

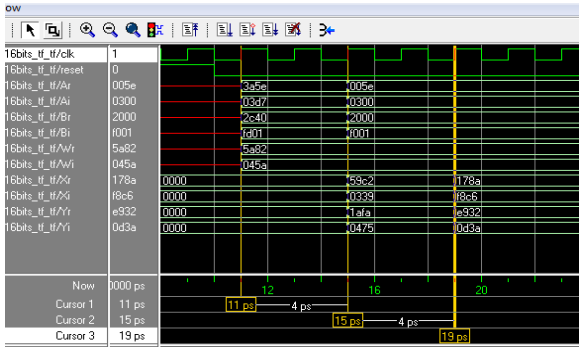

Fig. 13: Simulation result of Digit-slicing Butterfly.

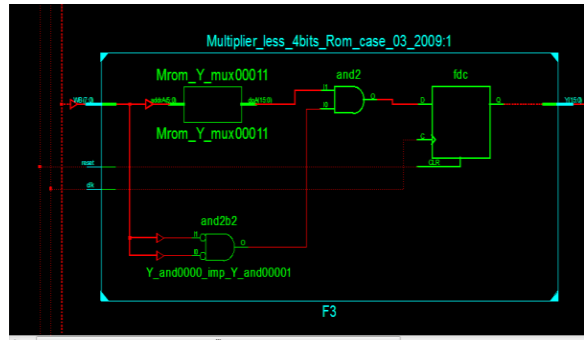

Fig. 15: RTL schematic for the Pipeline Digit-slicing Single Constant Multiplier-Less Lookup table (ROM) for the Butterfly.

The other module uses the pipelined digit-slicing single constant multiplier-less architecture to perform the multiplication with the twiddle factor. Both modules were built and tested in MATLAB as indicated in Fig. 9 and 10, and is then coded in Verilog and synthesized by using the XST-Xilinx Synthesis Technology tool. The target FPGA was Xilinx Virtex-II XC2V500-6-FG456 FPGA. The ModelSim simulation result of pipelined digit-slicing multiplier-less radix-2 DIT butterfly is shown in Fig. 12 and 13, while the synthesis results for the two models are presented in Table 1, which demonstrates the hardware specifications for the design. It indicates the maximum clock frequency of 549.75 MHz for Pipelined digit-slicing Multiplier-less Butterfly as well as the Pipelined Digit-slicing Single Constant Multiplier-less for the butterfly with a performance of the maximum clock frequency of 609.980 MHz. Meanwhile, Fig. 14 and 15 shows the RTL schematic for the Pipeline Digit-Slicing Single Constant Multiplier-less for the Butterfly.

Table 1: Hardware specifications of the digit-slicing butterfly

| Xilinx Virtax-II FPGA XC2v250-6FG456 | Total equivalent gate count for design | Maximum Frequency MHz |
|---|---|---|
| Conventional butterfly | 18.408 | 198.987 |
| Pipeline Digit-Slicing Multiplier-less Butterfly | 31.159 | 549.750 |
| Conventional 16 bits Multiplier | 4.131 | 220.160 |
| Pipeline Digit-Slicing Single Constant Multiplier-Less 16 bits for the butterfly | 6.483 | 609.980 |

## CONCLUSION

This study presented an on-chip implementation of pipeline digit-slicing multiplier-less butterfly for FFT structure. The implementation has been coded in Verilog hardware descriptive language and was tested on Xilinx Virtex-I1 XC2V500-6- FG456 prototyping FPGA board. A maximum clock frequency of 549.75 MHz has been obtained from the synthesis report for the pipeline digit-slicing multiplier-less butterfly that is 2.77 time faster than the conventional butterfly. It can be concluded that on-chip implementation of pipeline digit-slicing multiplier-less butterfly for FFT structure is an enabler in solving problems that affect communications capability in FFT and possesses huge potentials for future related works and research areas.